\documentclass[12pt,doublespace]{article}
\usepackage {amsmath}
\usepackage[dvips]{graphicx}
\usepackage{feynmp}
\usepackage{amssymb}
\usepackage{cite}
\newcommand{\be}{\begin{equation}}
\newcommand{\ee}{\end{equation}}
\newcommand{\bear}{\begin{eqnarray}}
\newcommand{\eear}{\end{eqnarray}}

\newcommand{\lapproxeq}{\lower .7ex\hbox{$\;\stackrel{\textstyle  
<}{\sim}\;$}} 
\newcommand{\gapproxeq}{\lower .7ex\hbox{$\;\stackrel{\textstyle  
>}{\sim}\;$}} 
\newcommand{\stackdown}[2]{\lower 1.4ex\hbox{$\;\stackrel{\textstyle{#1}}  
{\scriptstyle{#2}}\;$}}
\newcommand{\beq}{\begin{equation}} 
\newcommand{\eeq}{\end{equation}} 

\newcommand{\ba}{\begin{eqnarray}}
\newcommand{\ea}{\end{eqnarray}}

\newcommand{\bea}{\begin{eqnarray}}
\newcommand{\eea}{\end{eqnarray}}

\makeatletter 
\def\slash{\@ifnextchar[{\fmsl@sh}{\fmsl@sh[0mu]}} 
\def\fmsl@sh[#1]#2{%
  \mathchoice 
    {\@fmsl@sh\displaystyle{#1}{#2}}%
    {\@fmsl@sh\textstyle{#1}{#2}}%
    {\@fmsl@sh\scriptstyle{#1}{#2}}%
    {\@fmsl@sh\scriptscriptstyle{#1}{#2}}} 
\def\@fmsl@sh#1#2#3{\m@th\ooalign{$\hfil#1\mkern#2/\hfil$\crcr$#1#3$}} 
\makeatother 
\begin{document}
\newpage 
\begin{titlepage}  
\begin{flushright} 
\parbox{4.6cm}{UA-NPPS/BSM-10/01 }
\end{flushright} 
\vspace*{5mm} 
\begin{center} 
{\large{\textbf {The soft supersymmetry breaking in D=5 supergravity 
compactified on $S_1/Z_2$ orbifolds.
}}}\\
\vspace{14mm} 
{\bf G. A.~\ Diamandis}, \, {\bf B. C.~\ Georgalas}, \, 
{\bf P.~\ Kouroumalou } and \\ {\bf A. B.~\ Lahanas}
{\footnote{email alahanas@phys.uoa.gr}}
\vspace*{6mm} \\
  {\it University of Athens, Physics Department,  
Nuclear and Particle Physics Section,\\  
GR--15771  Athens, Greece}

\end{center} 
\vspace*{25mm} 
\begin{abstract}
 We study the origin of the supersymmetry breaking induced by the  mediation of gravity and the radion multiplet from the hidden to the visible brane  in the  context of the $N=2$, $D=5$ supergravity compactified on $S_1/Z_2$ orbifolds. The soft supersymmetry breaking terms for scalar masses, trilinear scalar couplings and gaugino masses are calculated to leading order in the five-dimensional Newton's constant $k_5^2$  and the gravitino mass $m_{3/2}$. 
These are finite and non-vanishing, with the scalar soft masses being non-tachyonic,  and are all expressed in terms of the gravitino mass and the length scale $R$ of the fifth dimension. The soft supersymmetry breaking parameters  are  thus correlated and the  phenomenological implications are discussed. 
\end{abstract}

\vspace*{2cm}
\end{titlepage} 
\newpage 
\baselineskip=17pt 
\section{Introduction}
The idea that our world is a brane embedded in a higher-dimensional space-time  has attracted much attention over the last decade, mainly because it offers new insights in particle physics beyond the standard model. It has opened new ways to resolve long-standing problems, and has brought new revolutionary concepts in Cosmology. 
Brane world models have been invoked for the hierarchy problem  \cite{randall1, randall2} as  an alternative towards explaining the large hierarchies between the Planck and the electroweak energy scales,  \cite{large1, large2, arkani1}.
Models with extra dimensions have  their origin in String Theory, where Supersymmetry is a basic ingredient \cite{horava, lykken, ovrut, smyth}. 
The mechanism of the supersymmetry breaking and the determination of the soft-breaking terms of the effective  low energy four-dimensional theory, is of outmost importance, especially in view of the LHC operation \cite{kane}. 
These models may be constructed by orbifolding a supersymmetric five-dimensional theory, and the supersymmetry breaking is triggered on the hidden brane which through the bulk is communicated to the visible brane, \cite{hiddensec,peskin, bagger1, nilles, fived, lalak, moroi, radion1, gates, riotto, anomalymed1, anomalymed2}. 
The transmission is done via the exchange of bulk gravitational fields, notably the radion multiplet, which are the messengers of supersymmetry breaking in this scheme. The induced corrections, which are finite, have been calculated \cite{riotto, rattazzi,scrucca, gregoire, tricher}, and  the resulting soft scalar masses on the visible brane were found to be tachyonic, although it has been clearly stated  that a proper treatment of the radion multiplet may turn this picture yielding positive masses squared. In Ref. \cite{peggy1,peggy2} these corrections were reconsidered, the soft scalar masses were found to be  non-tachyonic \cite{peggy2} and the induced soft scalar trilinear couplings were derived.
  
In this note, and in order to have a complete picture of the supersymmetry breaking sector, 
we study the transmission of the supersymmetry breaking in  $N=2$, $D=5$ supergravity \cite{fivesugra, gunaydin, recent} compactified on an $S_1/Z_2$ orbifold, in the on-shell scheme, and compute, in addition, the induced soft gaugino mass corrections when the theory possesses a gauge symmetry with gauge fields confined on the brane. 
This paper is organized as follows : \\ 
We first discuss the bulk-brane interactions, that induce soft supersymmetry breaking terms, and we calculate these to leading approximation in the five - dimensional gravitational coupling and the zero mode gravitino mass which sets the scale of supersymmetry breaking. They depend on the size of the $S_1/Z_2$ orbifold and the gravitino mass and are thus correlated. These can be used in the constrained MSSM to derive the mass spectrum which due to these correlations has distinct features from the popular MSSM schemes studied in literature. In particular the gaugino masses are negative, the trilinear soft scalar couplings are linearly dependent on the gaugino masses and the gravitino mass may be substantially larger than the soft scalar and gaugino masses and therefore the difficulties associated with Big Bang Nucleosynthesis can be, in principle, evaded. 
\section{Brane Couplings}

In the context of five-dimensional N=2 Supergavity orbifolds we addressed the problem of the coupling of N=1 multiplets, which live on the boundary branes, and we derived the interactions of the brane fields with  the bulk gravitational fields, \cite{peggy1,peggy2}. 
We found that the $N=1$ supersymmetric couplings of the brane chiral multiplets with the bulk fields  are determined by a K\"{a}hler function reminiscent of the no-scale model \cite{noscale,okada}. The Lagrangian derived in this way describes the full brane-radion coupling at least to order $k_5^2$, which is adequate for the derivation of the soft supersymmetry breaking terms to leading order, induced by the mediation of the  radion multiplet. 
For the derivation we worked in the on-shell scheme, avoiding the numerous auxiliary fields of the off-shell formulation. Besides in the on-shell scheme all possible gaugings have been classified which is essential for a unified description in which the Standard Model may be embedded \cite{gunaydin}.  

The radion multiplet 
propagates in the bulk but it also consists of even fields able to couple to the brane fields and therefore it can communicate supersymmetry breaking from one brane to the other. 
The scalar and the fermion fields of the radion multiplet are 
$$
T \equiv \frac{1}{\sqrt{2}} \;e_5^{\dot{5}} - \frac{i}{\sqrt {3}} \;A_5^0  \, \, , \, \, \, 
\chi ^{(T)} \equiv -\psi _5^2
$$
where $ A_\mu^0 $ is the graviphoton field, and the couplings of the brane fields are those of the N=1, D=4 supergravity derived from the following generalized K\"{a}hler function
\begin{eqnarray}
{\cal{F}}\;= -3\;ln\frac{T+T^*}{\sqrt{2}} + \delta (x^5)\;  \frac{\sqrt{2}}{T+T^*}\;K(\varphi ,\varphi ^*)\, ,
\label{kahler}
\end{eqnarray}
in the basis where the restriction of the Einstein-Hilbert part of the action on the brane is in its canonical form, for details see \cite{peggy1,peggy2}. 
In eq. (\ref{kahler}) the function  $K(\varphi ,\varphi ^*)$ depends on the scalar fields of the chiral multiplets living on the visible brane at $x^5 = 0$
{\footnote{
Our viewpoint is that in the absence of gravitational effects the brane action has the structure of a general $\sigma$ - model with a K\"{a}hler function given by $\,K(\varphi ,\varphi ^*)\;$. The delta function in front of $\,K(\varphi ,\varphi ^*)\;$ drops upon integrating the fifth dimension and is also encountered in the treatment of \cite{rattazzi}.
}}
. 
 
Introducing a superpotential $W(\varphi)$, involving brane fields, gives rise 
to the following Yukawa and potential terms \cite{peggy1},  
\begin{eqnarray}
\mathcal{L}_Y + \mathcal{L}_P 
&=& - e^{(4)} \Delta_{(5)}  e^{ \;\mathcal{F}/2} ( \;
W^* \psi_{\mu} \sigma ^{\mu \nu} \psi_{\nu}  
+ \frac{i}{\sqrt{2}} D_i W \chi^i \sigma^{\mu} \bar{\psi}_{\mu} 
+ \frac{1}{2} D_i D_j W \chi^{i}\chi^{j} + h.c. \;) \nonumber \\
&&- e^{(4)} (\Delta_{(5)})^2 e^{\;\mathcal{F}} (\;  
\mathcal{F}^{ij^*}\; D_i W \; D_{j^*}W^*  - 3  \; | W |^2 \;) \, \, \, .
\label{potential}
\end{eqnarray}
\noindent
In this, and in what follows, $\psi_{\mu}$ stands for $\psi_{\mu}^{1}$, the even gravitino field, which lives on the visible brane, and the bulk as well, and the function $\Delta_{(5)}$ is defined as 
$\Delta_{(5)} \equiv e^5_{\dot{5}} \, \delta (x^5) \,$. 

Without loss of generality, by a proper rescaling of the metric, the background geometry towards the fifth direction can be taken flat. However the metric on the 4-D brane can be curved in general. 
In any 4-D metric the arising scalar potential in this theory is positive definite. In particular  if a cubic superpotential is assumed for the visible sector the scalar potential involves  $\, {| \varphi |}^4 \, ,\, {| \varphi |}^6    \,$ terms, with positive coefficients, as has been pointed out in \cite{peggy2}. This is due to the no-scale form of the first term in the  K\"{a}hler function of eq. (\ref{kahler}). 
As a result no-cosmological constant is generated since the observable fields develop no vev's after spontaneous breaking of supersymmetry, which takes place on the hidden brane. A cosmological constant  arises after gauge symmetry breaking occurs. 

For the study of the induced supersymmetry breaking terms we need only pick the vertices that couple a bilinear of the gravitino, or the radion fermion, to the brane fields. In particular to 
$\phi^* \, \phi  $, for the calculation of the soft scalar masses, or to the superpotential $W(\phi)$ for the calculation of the soft trilinear scalar couplings. The relevant terms to that purpose are given by 
\begin{eqnarray}
\mathcal{L}
= - e^{(4)} \Delta_{(5)} \; ( \,
e^{ \;\mathcal{F}/2}  \;
W^* \psi_{\mu} \sigma ^{\mu \nu} \psi_{\nu}   
+ \;  \frac{ i }{2} \,K \,
\psi _{\dot{5}}^2\sigma^{\mu}D_{\mu} \bar{\psi} _{\dot{5}}^2  \, \;
) + h.c. \, \, .
\label{couplings2}
\end{eqnarray}
The last term in this equation yields the coupling of the radion multiplet fermion $ \psi _{{5}}^2 $ to $ \phi^* \phi  $
when $K$  is expanded as $K\,=\, \phi^* \phi + ...$ in order to yield canonical kinetic terms for the brane fields.
In the above formulae we have written the Lagrangian for a chiral multiplet located on the brane at $x^5=0$ 
but similar expressions hold for the hidden brane located at $x^5=\pi R$ as well. 
In this case we have just to add a hidden part  
$\mathcal{F}_H =  \delta (x^5-\pi R) \; \frac{\sqrt{2}}{T+T^*}\; K_H(\varphi_H ,\varphi_H ^*)$ 
 to the K\"{a}hler function ${\cal{F}}$ of eq. (\ref{kahler}), which  depends only on the hidden brane fields, and a corresponding hidden superpotential $W_H$. 

For the gauge multiplets confined on the visible brane, there are 4-fermion interaction terms coupling gauginos to gravitinos. By a Fierz rearrangement these that give rise to non-vanishing gaugino masses at one loop can be brought to the form,  
\be
- \frac{1}{4} e^{(4)} \delta (x^5) \; 
\left( \; Re f_{\alpha \beta} \;\bar{\lambda }^\alpha \bar{\lambda }^\beta  \psi _{\mu } \psi ^{\mu }  \,+\, h.c.\right) \,.
\label{gggrav}
\ee
In this $ \alpha, \beta  $ are gauge group indices and $ f_{\alpha \beta}( \phi ) $ is the gauge kinetic function which is a holomorphic function of the brane fields $\phi$. Other 
4-fermion terms, coupling $ \psi _{\mu } {\bar \psi}_{\nu }  $ to 
$ \lambda^\alpha {\bar \lambda }^\beta $, exist but do not induce through loops gaugino masses.

\section{Transmission of the Supersymmetry Breaking}
For the study of the transmission of supersymmetry breaking we consider 
a constant superpotential on the hidden brane. Other less trivial options lead to the same results concerning the forms of the induced soft SUSY breaking parameters, whose calculation is our main goal. The corresponding Lagrangian on the hidden brane can be read from the first term of eq. (\ref{couplings2}) translated for the hidden brane. 
Mass terms for the gravitino $\psi_{\mu}$ and the spinor field $\psi _{\dot{5}}^2$ of the radion multiplet arise only on the hidden brane while  non-minimal radion kinetic terms 
appear on the visible brane, as well, as can be seen from eq. (\ref{couplings2}). 
We shall follow a procedure in which the bulk kinetic terms of the gravitinos are disentangled from their fifth components and  mass terms are treated as vertex insertions which is sufficient for corrections to leading order in the gravitino mass  $m_{3/2} $.
We also take $K=\varphi \varphi ^*$  which is actually conceived as the leading term in the expansion of the K\"{a}hler function $K$ in inverse powers of the Planck mass. 

In our approach we employ the following gauge fixing,    
\begin{eqnarray}
i \; \frac{\xi }{2} \; \bar{\hat \Psi }_{i \tilde{m}}
 \gamma ^{\tilde{m}}  \gamma ^{\tilde{r}} 
 \gamma ^{\tilde{n}} \partial _{\tilde{r}} {\hat \Psi}_{\tilde{n}} ^i \; \;, 
 \label{gfix}
\end{eqnarray}
reminiscent of the one often used in the context of four-dimensional supergravity ( see for instance \cite{VanNieuwenhuizen:1981ae}),  which is added to the bulk Lagrangian 
{\footnote{ 
$ \; \; {\hat \Psi}_{\tilde{m}}^{1}= \left( \begin{array}{c} {\hat \psi}_{\tilde{m}}^{1}  \\ 
{\bar{{\hat \psi}}_{\tilde{m}}^{2} } \end{array} \right)$ and 
$ {\hat \Psi}_{\tilde{m}}^{2}= \left( \begin{array}{c} {\hat \psi}_{\tilde{m}}^{2}  \\ 
-{\bar{{\hat \psi}}_{\tilde{m}}^{1} } 
\end{array} \right)$ are symplectic Majorana gravitinos.
}}.
The gauge choice $\xi = - \frac{3}{4}$, and an appropriate shift of the fermions involved,  
eliminates the $m , \dot{5}$ kinetic mixings  \cite{peggy2} and 
the remaining terms that are left over, mixing 1 and 2, may  be treated conveniently by using the following Dirac spinors, 
\[
 \Psi = \left( \begin{array}{c} \psi _{\dot{5}}^{2 }  \\\bar{ \psi} _{\dot{5}}^{1 }  \end{array} \right)\,\,,
\,\, \Psi_m = \left( \begin{array}{c} \psi _m ^{1 }  \\\bar{ \psi} _{m}^{2 }  \end{array} \right) \; \; .
\]

For the calculation of the induced soft SUSY breaking terms, we need study diagrams involving propagations from the hidden to the visible brane and we use the pertinent Dirac and gravitino propagators in the mixed momentum-configuration space representation \cite{arkani1, puchwein, meissner}. In this representation, and in the particular gauge fixing, with the value of  $\xi$ chosen as above, the orbifolded propagators read as \cite{peggy2},
\begin{eqnarray}
&&G_{mn}(p,y,y') = \left( \frac{1}{2} \gamma _n \slash{p} \gamma _m +
 i \eta _{mn} \gamma ^{\dot{5}} \partial _y \right) \; F(p,y,y') \nonumber \\
&&G (p,y,y') = \frac{2 \;i}{9} ( \slash{p} + i  \gamma ^{\dot{5}} \partial _y ) \; F(p,y,y') 
\end{eqnarray}
with the function  $F(p,y,y') $ given by 
\begin{eqnarray}
&&F(p,y,y') \equiv  \frac{1}{2 \; q \;sin(q \pi R)}
 \left\{
cos \left[q(\pi R - \mid y - y' \mid )\right] - i \gamma ^{\dot{5}} cos \left[q(\pi R -  y - y') \right] 
  \right \} \;  .
\nonumber
\end{eqnarray}
In these $y, y'$ denote variables along the fifth dimension and $q = \sqrt{-p^2 + i\epsilon }$. 

Note that the fermions $ \, \Psi, \Psi_m \, $ carry dimension two, i.e. their mass dimension  in the five-dimensional space-time and for 
either brane their  couplings to superpotential terms are given by 
\begin{eqnarray}
- \,k_5^2 \,e^{(4)} \, {\cal{W}} \;  \left[ \bar{\Psi}_m  ( \sigma^{mn} -\frac{1}{3}\gamma ^{m} \gamma ^{n} )  
C P_R \bar{\Psi}_n^T  + 
\frac{3}{2} \Psi ^T  C P_L \Psi - \frac{i}{2} \bar{\Psi }_m  \gamma ^m P_L \Psi  \right] + h.c. 
\; \; 
\label{super}
\end{eqnarray}
where $\, {\cal{W}} \,$ is the superpotential on the corresponding brane
{\footnote{
${\cal{W}}  \,$ denotes the superpotential of the observable fields $\,W(\Phi) \, $, if we are on the visible brane, or that of the hidden fields $\,W_H \, $ if we are on the hidden brane. In eq. (\ref{super})
$  \sigma^{mn} = \frac{1}{4} \, \left[ \gamma^m , \gamma^n   \right] $.
}}
. 
In this equation 
we have reinstated dimensions by introducing the coupling $ k_5^2 $ of the five-dimensional gravity. 
Also for convenience, in this and the following equations, we neglect the delta functions that accompany the brane Lagrangian densities. 
 Throughout  $P_{L,R} \;$  stand for the chiral projection operators and $C$ denotes the charge conjugation matrix. 

The mass terms on the hidden brane that are the sources of supersymmetry breaking follow by 
considering a constant superpotential $\, {\cal{W}} = c \,$. 
The supersymmetry breaking scale is then set by 
\be
m_{3/2} \,=\, k_5^2 \; \dfrac{ \mid c\mid}{\pi \, R} \; \; .
\label{gravmass}
\ee
which is actually the mass of the zero mode gravitino, \cite{bagger1,riotto}.
 
\begin{figure}
\begin{center}
\includegraphics[width=13cm]{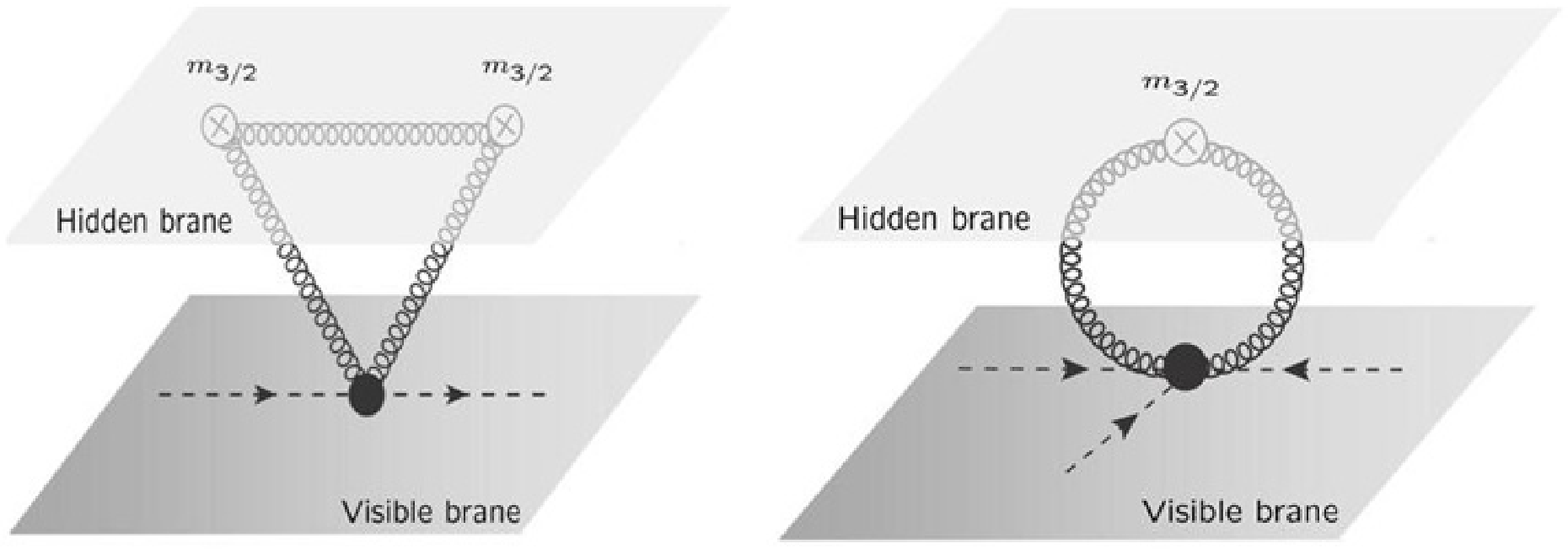}
\end{center}
\caption[]{Diagrams relevant for the calculation of the induced scalar field masses (left) and trilinear scalar couplings  (right). The dashed lines denote the scalar fields living on the visible brane. The curly lines denote either gravitinos or spinor fields of the radion multiplet. The blobs  are fermionic mass insertions on the hidden brane.}
\label{fig1}  
\end{figure}

Starting the discussion with the soft scalar masses, the pertinent interaction terms on the visible brane coupling $ \varphi \varphi ^* $ to pairs of  $ \Psi $, $ { \Psi}_m $ fermions are of the form 
\[
-i \,k_5^2\, e^{(4)} \,\varphi \varphi ^* \left[ \bar{\Psi} \slash{\partial } P_L \Psi  + 
\frac{1}{9} \bar{\Psi}_m \gamma ^m  \slash{\partial } \gamma ^n P_L \Psi _n +  \frac{i}{3}(
 \Psi_m ^T  C \gamma ^m \slash{\partial } P_L \Psi  - h.c )\right] \; .
\]   
Calculating the  diagrams  depicted on the left pane of Fig. \ref{fig1},  
we find the mass corrections to the scalar fields involved. 
The momenta of the external scalar fields in these graphs have been taken vanishing.
The general structure of the loop involved is 
\begin{eqnarray} 
\int \frac{d^4p}{{(2 \pi)}^4}  \;
Tr \left[  V G(p, 0,\pi R) V_1 G(p,\pi R,\pi R) V_2 G(p,\pi R,0) \right] \; , \nonumber
\end{eqnarray}
where all space-time indices have been suppressed and $V$ and $ V_{1,2}$ are vertices on the visible and hidden brane respectively. $G(p,z,z')$ denote propagations between $z$ and $z'$ points for the gravitino and fermion fields carrying the loop momentum $p$. Integration over the variables $y_{1,2}$ specifying the points on the hidden brane, located at $\pi R$, and $y$ on the visible brane, located at $y=0$, have been performed, \cite{peggy2}.

Collecting all contributions entails to the following finite mass correction to lowest order in the 
gravitino mass 
\begin{eqnarray}
m_0^2 = \frac{\zeta (3)}{16 \pi^5} \; \frac{k_5^2}{R^3}\;   m_{3/2}^2 \; = 
\frac{\zeta (3)}{\pi^3 R^2} \; \frac{m_{3/2}^2}{M_{Planck}^2} \; .
\label{scalarsoft}
\end{eqnarray}
Note that the resulting scalar masses are non-tachyonic \cite{peggy2}. 
To this order in the gravitino mass, higher loops contribute $\sim  m_{3/2}^2 \; g^N  $ where $N$ is the loop order and $g$  the dimensionless expansion parameter
\begin{eqnarray}
g \;=\; \dfrac{1}{\pi^2} \; \dfrac{ k_5^2}{V_5^3 } \;.
\label{coupdim}
\end{eqnarray}
In eq. (\ref{coupdim}) $ V_5 $ is the "volume" $ 2 \pi R$ of the orbifold $S_1/Z_2$. The coupling $g$ oughts to be small  
$\;g \lesssim 1  $ for the calculation to be valid which entails to $ R^{-1} \; \lesssim  \; 4 \; M_{Planck} \;$.
Therefore large values of the radius $R$ are acceptable in this scheme. 
 
For the study of the effect of the supersymmetry breaking on the trilinear scalar couplings we consider a cubic superpotential on the visible brane $W(\Phi ) = \dfrac{\lambda }{6}\; \Phi ^3$. The graphs that need be considered for the induced trilinear couplings are shown on the right pane of Fig. \ref{fig1} and 
the pertinent Yukawa-type Lagrangian terms for this computation, to leading order in $ k_5  $, are read from  
Eq. \ref{super}  with $\;{\cal{W}}\;$ replaced by $W(\phi ) \;$ above.
Then the corrections to the superpotential, due to the supersymmetry breaking triggered on the hidden brane, can be calculated and the induced trilinear soft scalar coupling is \cite{peggy2},
\be
A_0 \;=\; 
3 \; \frac{\zeta (3)}{16 \pi^5} \; \frac{k_5^2}{R^3}\;   m_{3/2} \;=\; 
3 \; \frac{\zeta (3)}{\pi^3 R^2} \; \frac{m_{3/2}}{M_{Planck}^2} \; .
\label{asoft}
\ee

\begin{figure}
\begin{center}
\includegraphics[width=13cm]{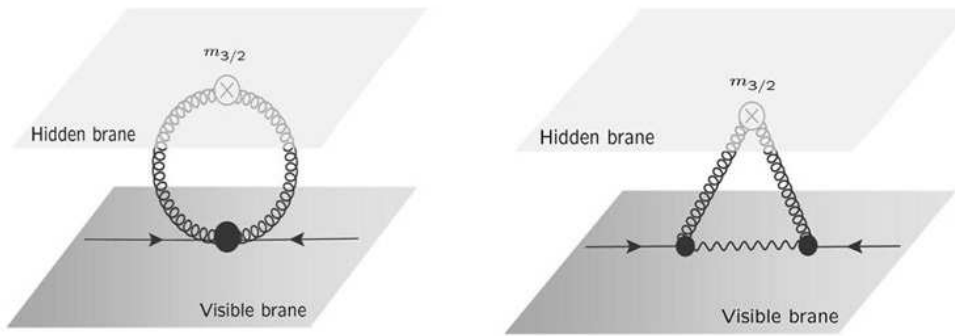}
\end{center}
\caption[]{Diagrams resulting to non-vanishing soft gaugino masses. The curly lines are as in Fig. \ref{fig1}. 
The solid lines represent gauginos and the wavy lines gauge bosons.}
\label{fig3}  
\end{figure}

For the gaugino mass corrections, adopting the four-component notation we have used previously, we find from 
eq. (\ref{gggrav}) the four-fermion interaction 
\begin{equation}
\Delta \mathcal{L}_{vis} \,=\, - \kappa _5^2 \, \frac{ e^{(4)}}{4} 
 Re f_{\alpha \beta} \; \left(\, {\bar \lambda}_R^\alpha \lambda_L^\beta \; \bar{\Psi}^{m } C P_R \bar{\Psi}_{m}^{T} \, \right)
\, +\, h.c. \,, 
\end{equation}
which induces a gaugino mass through the graph  shown on the left pane of Fig. \ref{fig3}. However there is an additional  interaction term
\be
\Delta {\mathcal{L}'}_{vis} \,=\,  \frac{i}{2}     \, \kappa _5 \, {e^{(4)}} \,
Re f_{\alpha \beta} \;
\left(
\,\bar{\Psi}_{m } \sigma^{rs} \gamma^m \, \lambda_L^\alpha 
 \, F_{rs}^\beta \right) \, + h.c. 
\ee
giving rise to the graph shown on the right pane of the same figure. 
The induced gaugino mass correction from the two graphs is found to be 
\be
 m_{1/2 } \,=\, 
- \frac{7}{2} \; \frac{\zeta (3)}{16 \pi^5} \; \frac{k_5^2}{R^3}\;   m_{3/2} \;=\; 
- \frac{7}{2} \; \frac{\zeta (3)}{\pi^3 R^2} \; \frac{m_{3/2}}{M_{Planck}^2} \; .
\label{gaugesoft}
\ee
which is non-vanishing
{\footnote{
We limit ourselves to cases where $ \dfrac{\partial f_{\alpha \beta}}{\partial \phi_i }$ does not develop a vacuum expectation value, due to appearance of high energy scalar fields, and therefore (\ref{gaugesoft}) is the only contribution to the gaugino masses. 
}}
.
The total coefficient $7/2$ is  the sum of $2$ and $3/2$ which are the separate  contributions of the left and right graphs displayed in  Fig. \ref{fig3}. Their contributions do not cancel, unlike  the case of 4-D supergravity 
\cite{barbieri} and the findings of \cite{riotto} in the case of 5-D brane models. We remark  that in our work the gravitino bilinear terms on the hidden brane, and hence the gravitino propagator, are different since by a field redefinition we disentangled the four-dimensional gravitino from the radion fermion $\, \psi_5^2 \;$. 
Also, our approach is quite different from that employed in other works where the gauge 
$\,  \psi_5^2 \; = \;0 \,$ is chosen to eliminate the radion fermion. However since the use of this gauge can be implemented only conditionally \cite{bagger1}, the vanishing of $\,\psi_5^2 \; $ over all five-dimensional space-time is rather questionable \cite{belyaev}. This is also supported by the appearance of pseudo-Goldstone fermion modes in the fermionic mass spectrum, \cite{benakli}. This is the reason we employed a different gauge fixing given by eq. (\ref{gfix}). In this gauge non-vanishing interactions of the $\,\psi_5^2 \; $ with the brane fields are present which play an important role, as we have seen, in the calculation of the soft SUSY-breaking  parameters.  
 
The effective theory on the visible brane is an ordinary 4-D supergravity  with soft SUSY breaking terms, which besides the observable fields, that are located only on the visible brane, also includes the projection of the radion multiplet on this brane which is an even member of the 5-D gravity multiplet. This is a generic feature in this class of models. The soft terms arise after spontaneous supersymmetry breaking which occurs on the hidden brane and is communicated to the visible sector through the bulk gravitational interactions. This picture is different from ordinary 4-D models where supersymmetry breaking takes place in a  hidden sector, which lies on the same brane along with the observable fields. Besides the soft scalar masses, which are non-tachyonic, and the calculation of the soft trilinear couplings,  an  essential element of our approach is the appearance of induced gaugino masses which  are proportional to the zero mode gravitino mass $\, \sim \, m_{3/2} \,$. These are due to the interactions of the radion-fermion as explained in detail above. 

The same mechanism that produces soft SUSY breaking terms does not shed any light on either the origin or the magnitude of the $\mu$ - term, at least at this loop order. Having in mind that the effective theory is a 4-D supergravity the major mechanisms invoked for a resolution of the $\mu$ - problem are applicable in this case as well.

\section{The phenomenology of the soft SUSY breaking}
The induced soft supersymmetry breaking parameters are correlated, as can be seen from 
Eqs. \ref{scalarsoft}, \ref{asoft} and \ref{gaugesoft}, since  
they are all expressed in terms of the gravitino mass $ m_{3/2} $ and the "volume" of the fifth dimension set by the radius  $ R \;$. For phenomenological purposes, one can treat as parameters the soft scalar and  gaugino masses,  
$ m_{0} $ and $m_{1/2}$ respectively, and then the trilinear scalar coupling $ A_0 $, as well as the gravitino mass $m_{3/2}$ and the radius $ R \;$, are determined. We have in mind a scenario which mimics the four-dimensional  mSUGRA with universal boundary conditions at the unification scale, defined as the point where the gauge couplings $\alpha_1$ and  $\alpha_2$ meet. 

The value of the common gaugino mass turns out to be negative 
{\footnote{
In the convention in which the gaugino terms are $ \, - \frac{M}{2} \, \lambda \lambda + h.c.  \, $ and the gaugino fields  appear in the vector multiplet as 
$\, V = ... +i \, \theta \theta \bar{\theta} \bar{\lambda} - i \bar{\theta} \bar{\theta} \theta \lambda  \, $. 
This  convention is followed by many authors.
}}
with magnitude given by
$$
 \mid m_{1/2 } \mid \;
\simeq \; 0.136 \; \; \frac{m_{3/2}}{R^2 \; M_{Planck}^2} \; \, .
$$
If the gravitino mass is in the TeV range so is $ m_{1/2 } $, provided the inverse radius $ R^{-1} $ does not lie far from the Planck mass, $ R^{-1} \simeq 3 \, M_{Planck} $. This is within the limits set by $g \lesssim 1$ ( see eq. (\ref{coupdim}) and discussion following it ). 
Much lower values of the gravitino mass result to values of $ R^{-1} $ considerably higher than the Planck mass, if the gaugino mass $ m_{1/2 } $  is in the TeV range, and hence unacceptable if the model is to be considered as an effective theory of the superstring. On the other hand, values of the gravitino mass larger than about $10\;$ TeV, as demanded by Big Bang Nucleosynthesis, result to   
$ R^{-1} \lesssim 0.857 \, M_{Planck} \sqrt{ {\mid m_{1/2} \mid }  \; / \; TeV   }$  and hence  the gravitino crisis may be evaded keeping at the same time the gaugino mass in the TeV range and $ R^{-1}$ within acceptable limits. 

In terms of $ m_{3/2} $ and $m_{1/2}$ the common soft scalar masses and the trilinear scalar couplings are
\be
m_0 \;=\; \sqrt{  \dfrac{2\, \mid m_{1/2 } \mid \; m_{3/2}  }{7}   } \quad, \quad 
A_0 \;= \; \dfrac{6}{7} \; \mid m_{1/2 } \mid \quad .
\label{aaa}
\ee
We observe that the soft trilinear coupling $A_0$ is always positive and non-vanishing and grows linearly with increasing the gaugino mass,  while the soft scalar mass grows with the square root of the gaugino mass, if the gravitino mass is kept fixed. 
Since the common gaugino mass is bounded from below, in order to comply with the lower experimental bound put on the chargino mass and/or absence of electroweak symmetry breaking, the second of eq. (\ref{aaa})  puts a lower bound on the trilinear coupling roughly given by $A_0 > 150 \, GeV \,$. This, along with the fact that it becomes large for large values of the gaugino mass, results to a mass spectrum  which differs from that obtained in the popular supersymmetric models where the trilinear coupling is allowed to be fixed, or small in comparison to the other soft SUSY breaking parameters involved.
\begin{table}
\begin{center}
\begin{tabular}[H!]{ccc} 
{\bf{Inputs: }} 
   &  $ {\mathbf{\quad m_0 \, , \,  m_{1/2} \, = \, 1000 \, , \, - 1550\, GeV   }}$  &  $ {\mathbf{ \quad tan \beta = 20 }}$
   \\  [3pt]
\hline
{\rule{0pt}{2.8ex}}
 mass (in GeV)  &   \quad \quad  This model: $A_0  =1329.6 \,GeV$ &  \quad \quad mSUGRA: $A_0 = 0 \,GeV$ 
\\  [2pt]
\hline\hline
{\rule{0pt}{2.5ex}}
$\mu(Q)  $    & $- 1523.0$ & $- 1672.9$ \\ 
$ \tilde{b}_{1,2} $    & 2868.6 , 2931.5 & 2839.2 , 2919.4  \\ 
$ \tilde{t}_{1,2} $    & 2482.2 , 2876.8 & 2419.6 , 2847.4  \\ 
$ {\tilde{\chi}}^+ $   & 1219.6 , 1533.1 & 1224.8 , 1680.1  \\ [2pt]
\hline
{\rule{0pt}{2.5ex}}
$ {\tilde{\chi}}^0 $   & \, \, 677.1 , \,1219.6  & \, \, 678.3 , \, 1224.8  \\ 
   & - 1524.7 , 1532.9    &  - 1674.6 , 1679.8    \\ [2pt]
\hline
{\rule{0pt}{2.5ex}}
$ h_0, H_0$   & \, 121.1 , 1956.4  & \, 122.0 , 2047.2  \\ 
$H_A, H_+$  &  \, 1956.4 , 1957.9   &  \,  2047.2 , 2048.5  \\
\hline
\end{tabular}
\caption{Predictions for the model presented in this work and in mSUGRA with $A_0 = 0$. The masses of the remaining sparticles are almost identical in the two models and are not shown. The mass spectrum has been derived using Suspect2, \cite{suspect} with $\mu < 0$. }
\label{tablechar}
\end{center}
\end{table}
 
In Table \ref{tablechar}, for comparison, we display the predictions of the model presented in this paper and mSUGRA 
for the inputs shown on the top of the table, 
$ \, m_0 \, , \,  m_{1/2} \, = \, 1000 \, , \, - 1550\, GeV   $, $ \quad tan \beta = 20 $. For the top the most recent experimental value for its pole mass has been used,  $ m_t = 173.1 \, GeV \,$ \cite{top}. The input running bottom mass is taken $ m_b(m_b) = 4.25 \, GeV $. In the mSUGRA case  the common trilinear scalar coupling is taken  $A_0 = 0 \,GeV$ while in the model discussed in this work the trilinear coupling is constrained to be 
$A_0  = 1329.6 \,GeV \,$, for the particular choice of the parameters $ \, m_0 \, , \,  m_{1/2} \,  $. 
This difference in the value of $A_0 \,$ affects the predicted value of the $\mu$-parameter at the average stop mass scale Q, by as much as 10 \%, affecting in turn the mass spectrum and in particular the heavy neutralino and chargino states. 
In the Table  \ref{tablechar} we only display the masses of the sectors that are affected most. Due to the 10 \%, difference in the predicted value of $\mu$ the heavy chargino and neutralino states differ by the same amount, while the effect is less, $\simeq$ 5 \%,  in the heavy Higgses. For the stops and sbottoms, their masses are shifted towards  higher values by as much as ${\cal{O}}(1\%) \;$, with their mass splittings decreasing, 
in the model under consideration, relative to the mSUGRA predictions.  This is due to the fact that both trilinear scalar couplings and  the values of $\mu$ at low scales, that control the L and R-handed sfermion mass mixings, are different in the two cases under consideration. 
For the remaining sectors the differences in masses are imperceptibly small. 
 
\begin{figure}
\begin{center}
\includegraphics[width=13cm]{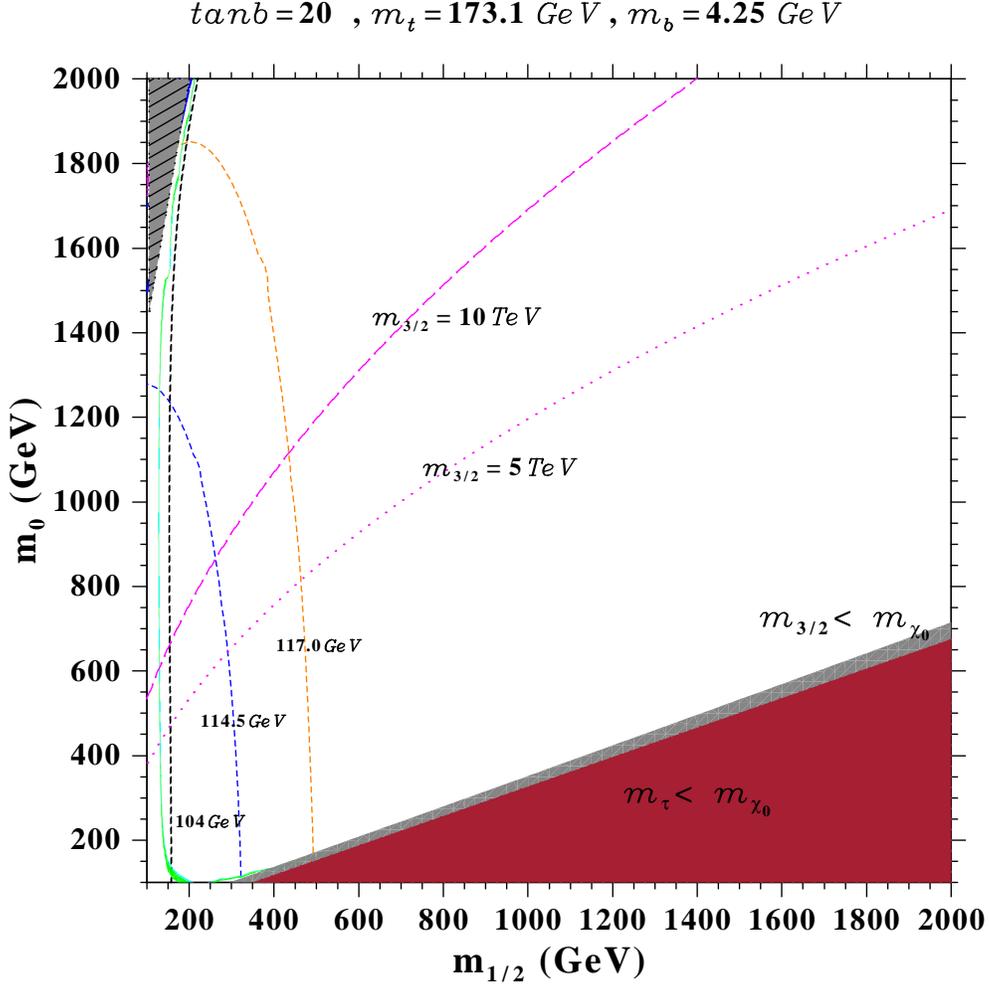}
\end{center}
\caption[]{
The $m_0, m_{1/2}$ plane for the soft supersymmetry breaking parameters  discussed in the main text for the input values shown on the top. The green thin region is cosmologically allowed by the WMAP data. In the grey thin stripe, just above the $ \, m_{\tilde \tau} < m_{\chi_0}  \, $ region, the gravitino is the LSP. The lines $ m_{3/2}=10 \; TeV $  and  $ m_{3/2}=5\; TeV $ have been drawn as well as the lines on which $m_{Higgs}$ is  $114.5 \; GeV $  and $117 \; GeV $.  The chargino mass bound is designated by the $104\, GeV$ line. In the shaded pattern there is no-electroweak symmetry breaking. 
The $b \longrightarrow s \,+\, \gamma $ and the $g-2$ regions have not been drawn. 
}
\label{fig4}  
\end{figure}

In Fig. \ref{fig4}, and for the particular inputs shown, we delineate, in the $ \,m_0, m_{1/2} \,$ plane,  the cosmologically region allowed by the WMAP data. The grey thin stripe that extends  just above the $ \, m_{\tilde \tau} < m_{\chi_0}  \, $ boundary, designates the region in which the gravitino is the LSP. Since the gravitino is the LSP in this region the bino-like coannihilation tale, which in the conventional schemes is usually extended up to values of $m_{1/2} \simeq 700 \, GeV \,$, is now excised. This is actually a generic feature of this class of models. 

For comparison the lines $ m_{3/2}=10 \; TeV $  and  $ m_{3/2}=5\; TeV $ have been drawn as well as the lines on which the Higgs masses $m_{Higgs}$ are  $114.5 \; GeV $  and $117 \; GeV $ respectively.  The chargino mass bound is designated by the dashed almost vertical line labelled by $104\, GeV$. In the shaded pattern there is no-electroweak symmetry breaking. The $b \longrightarrow s \,+\, \gamma $ and the $g-2$ regions have not been drawn.  
One observes that values of the gravitino heavier than about $\simeq 5 \, TeV$  
do not overlap with regions allowed by WMAP and other available data  
and it seems there is no way to reconcile experimental data with constraints put by Nucleosynthesis. However  gravitino masses of this magnitude occupy regions of the $ \, m_0 \, , \,  m_{1/2} \,$ plane in which  bino-like LSP's rapidly annihilate through a Higgs resonance, when  $\, \tan \beta > 45 \, $ is large, and extended funnel-like cosmologically allowed domains start to show up. Therefore is not unlikely that for large 
$\, \tan \beta \, $ WMAP data and heavy gravitino masses, as demanded by Nucleosynthesis,  may coexist while 
$ \, m_0 \, , \,  m_{1/2} \,$  are kept low within the reach of LHC. These issues are under consideration and will be presented in a forthcoming publication. 
\section{Conclusions}
In the context of the $D=5$, $N=2$, supergravity compactified on 
$S^1/Z_2$ we studied the transmission of the supersymmetry breaking occurring on the hidden brane to the visible sector of the theory. To leading order in the gravitino mass $m_{3/2}$ we considered the one loop soft-breaking parameters, induced by the supersymmetry breaking occurring on the hidden brane and transmitted to the visible brane by the propagation of the radion multiplet. 
We show that this transmission results to non-tachyonic universal masses $\;m_{0}^2 > 0 \;$ for the visible scalar fields and also non-vanishing gaugino masses. All soft SUSY breaking terms depend on two parameters, the gravitino mass and the "volume" of the fifth dimension and are thus  correlated. The soft trilinear scalar couplings are found to be  strictly positive, and non-vanishing, linearly dependent on the gaugino masses. The trilinear couplings become large, for large values of the gaugino mass, resulting to a mass spectrum whose heavy neutralino and chargino states have masses differing from that obtained in the popular mSUGRA models. 
Besides the gravitino is predicted to be the LSP in a region which slightly extends above the region in which the stau is lighter than the lightest neutralino. This results to a shrink of the cosmologically coannihilation allowed domain of the conventional supersymmetric models, in which the LSP is a bino-like neutralino with mass close to that of the stau sparticle. 
In this class of models large gravitino masses, $m_{3/2} > 5 \, TeV \, $ occupy regions of the $ \,m_0, m_{1/2}  \,$ plane in which rapid bino-like LSPs annihilations via a pseudoscalar Higgs resonance can take place for large values of $ \tan \beta $. It is not unfeasible therefore that regions allowed by WMAP and other accelerator data can conform with the constraints of Big Bang Nucleosynthesis. These issues are under study and will appear in a forthcoming publication. 

\section*{Acknowledgements}
A.B.L. acknowledges support  by the European Union through the fp6  Marie-Curie Research and Training Networks, UniverseNet (MRTN-CT-2006-035863) and Heptools (MRTN-CT-2006-035505). The authors wish to acknowledge partial support from  the University of Athens Special Research Account. 


\newpage

\end{document}